\documentclass{mem}
\usepackage{natbib}
\usepackage{txfonts}\usepackage{balance}
\usepackage{graphicx}
\usepackage[a4paper]{hyperref}
\idline{XX}{1}
\begin{document}

\newcommand{\text}[1]{\quad\mbox{#1}\quad}
\newcommand{\spr}[2]{\vec{#1} \!\cdot\! \vec{#2}}
\newcommand{\vpr}[2]{\vec{#1} \!\times\! \vec{#2}}
\newcommand{\vgrad}[1]{\vec{\nabla}{#1}}
\newcommand{\vdiv}[1]{\spr{\nabla}{#1}}
\newcommand{\vcurl}[1]{\vpr{\nabla}{#1}}
\newcommand{\Lop}{\triangle}
\newcommand{\Dop}{\Box}
\newcommand{\E}{{\cal E}}
\newcommand{\oder}[2]{\frac{d #1}{d #2}}
\newcommand{\pder}[2]{\frac{\partial #1}{\partial #2}}
\newcommand{\Pd}[1]{\partial_{#1}}
\newcommand{\Od}[1]{\frac{d}{d #1}}
\newcommand{\ort}[1]{ \vec{i}_{#1} }
\newcommand{\ortt}[1]{ \vec{i}_{\hat{#1}} }
\newcommand{\mean}[1]{ <\!#1\!> }
\newcommand{\fracb}[2]{\left(\frac{#1}{#2}\right)}

\title{Magnetic acceleration of relativistic jets}
\subtitle{}

\author{S.S. \,Komissarov\inst{}}

\institute{
School of Mathematics,
University of Leeds, 
Leeds LS29JT, UK
\email{serguei@maths.leeds.ac.uk}
}

\authorrunning{Komissarov }

\titlerunning{Magnetic acceleration}

\abstract{ This is a brief review of the recent developments in the theory 
of magnetic acceleration of relativistic jets. We attempt to explain the key results of this complex theory using basic physical arguments and 
simple calculations. The main focus is on the standard model, 
which describes steady-state axisymmetric ideal MHD flows. We argue that 
this model is over-restrictive and discuss various alternatives.  
}

\maketitle{}

\section{Introduction}

Collimated flows, or jets, are observed in a variety 
of astrophysical systems but the most spectacular examples are related to 
disk accretion onto a compact central object. This suggests that 
disk accretion is an essential element of cosmic jet engines, 
via providing a source of power or collimation, or both.
The most remarkable property of astrophysical jets is that their length 
exceeds the size of the compact object, and hence the size of the 
central engine, by many orders of magnitude. For example, AGN jets are 
generated on the scale no more than hundred gravitational radii of 
the central black hole, $\sim 10^{15}$cm,  and propagate up to 
the distance of $\sim 10^{24}$cm, where they create extended radio lobes. 
Along the jets, the specific volume of plasma increases enormously 
and the corresponding adiabatic losses, in combination with 
various radiative losses, ensure that the plasma particles loose 
essentially all their ``thermal'' energy, which they might have had inside 
the central engine, very quickly. Yet, the observations 
show that the jet brightness does not decline so rapidly. 
This suggests that most of the jet energy is in a different form and 
that the observed emission is the results of its slow dissipation.    

One possibility is that astrophysical jets are supersonic, kinetic 
energy-dominated  flows \citep{S74,BR74}. 
Indeed, a number of factors make the idea very attractive. 
First, such flows do not require external support in order to preserve their 
collimation. Second, they are much more stable and can propagate large 
distances without significant energy losses, in an essentially ballistic regime. 
Third, when they interact with the external medium the result is 
shocks, which dissipate kinetic energy locally and thus can produce 
bright compact emission sites, reminiscent of the knots and hot spots of 
astrophysical jets.    

Generic industrial jet engine consists of a chamber, which is being filled with 
very hot gas, and a carefully designed nozzle through which the gas escapes with 
a supersonic speed. The acceleration mechanism is thermal -- it is the thermodynamic 
pressure force that accelerates this gas and convert its thermal energy into the 
bulk motion kinetic energy. Apparently, a very similar configuration may form 
naturally during the gravitational collapse of rotating massive stars.  
It has been shown that, for a sufficiently rapid rotation, 
the centrifugal force eventually ``evacuates'' the 
polar region around the central black hole. The neutrinos, emitted from 
its super-critical accretion disk, can then fill the cavity with  
ultra-relativistically hot plasma via neutrino-antineutrino annihilation.         
This plasma can then expand in the direction of least resistance, which is 
obviously the polar direction, and thus create a collimating nozzle.    
In fact, this may the origin of jets associated with long GRBs \citep{MW99}.  

Similar ideas have been explored earlier for ANG jets but the results were 
rather discouraging, which  stimulated search for alternatives. 
Gradually, the magnetic model emerged, where the Maxwell stresses were responsible 
for powering, acceleration, and even collimation of jets. In principle, this 
model even allows to avoid the kinetic energy-dominated phase 
as the observed emission can be powered via dissipation of magnetic energy 
\citep[e.g.][]{LB03}. 
However, thanks to the success of the shock model in many areas 
of astrophysics, the current magnetic paradigm assumes that most of the jet
Poynting flux is first converted into bulk kinetic energy and then dissipated at 
shocks (shocks in highly magnetised plasma are inefficient).         

%sssssssssssssssssssssssssssssssssssssssssssss
\section{The Standard Model}
%sssssssssssssssssssssssssssssssssssssssssssss

The problem of magnetic acceleration is rich and mathematically complex. 
This complexity is the reason why most theoretical efforts have 
been directed towards development of the {\it standard model}, which deals with 
steady-state axisymmetric flows of perfect fluid. Although this is the simplest
case it is still impossible to give a comprehensive and mathematically rigorous 
review within the scope of this presentation. Instead, I will focus more on 
physical arguments.      

%sssssssssssssssssssssssssssssssssssssssssssss
\subsection{Acceleration in supersonic regime} 
%sssssssssssssssssssssssssssssssssssssssssssss

One clear difference between non-relativistic and relativistic flows is that 
in the non-relativistic limit most of the energy conversion occurs 
already in the subsonic regime whereas in the relativistic limit this 
occurs in the supersonic regime. This is true both for the thermal and 
magnetic mechanisms 

Consider first the non-relativistic limit. 
For a hot gas with polytropic equation of state, the thermal energy density 
$ e= p/(\gamma-1)$ and the sound speed $ a_s^2 = \gamma {p}/{\rho}$, 
where $\gamma$ is the ratio of specific heats, $p$ and $\rho$ are the gas 
pressure and density respectively. From this we immediately find that 
when the flow speed equals to the sound speed  
$\rho v^2 = \gamma(\gamma-1)e.$
Thus, the kinetic energy is already comparable with the thermal energy.
For a cold magnetised flow the speed of magnetic sound (fast magnetosonic 
wave) is $ c_f^2 = B^2/{4\pi \rho} $ and at the sonic point   
$\rho v^2 =  B^2/4\pi.$
Thus, the kinetic energy is already comparable with the magnetic energy. 

The relativistic expression for the sound speed is 
$ a_s^2 = (\gamma {p}/{w}) c^2 $, where 
$ w = \rho c^2 + {p\gamma}/{(\gamma-1)} $ is the gas enthalpy, 
$\rho$ and $p$ are measured in the fluid frame. 
The condition of highly-relativistic asymptotic speed requires 
relativistically hot gas (initially), that is the Lorentz factor 
of thermal motion $\Gamma_{\rm th} \gg 1$, $ p \gg \rho c^2$, and 
$\gamma\approx4/3$. This means that at the sonic point  
$v^2 \approx  c^2/3$ and the corresponding Lorentz factor is only 
$\Gamma^2 \approx 3/2 \ll \Gamma^2_{\rm th}$.  
Thus, most of the energy 
is still in the thermal form.  

The relativistic expression for the speed of magnetic sound in cold 
gas is 
\begin{equation} 
\label{eq:cf}
c_f^2=B'^2/(B'^2+4\pi\rho c^2),
\end{equation}
where $B'$ is the magnetic field as measured in the fluid frame. 
Thus, at the sonic point 
$\Gamma^2 = 1 + {B'^2}/{4\pi\rho c^2}$. Now one can see that 
large asymptotic Lorentz factor implies ${B'^2}/{4\pi\rho c^2}\gg1$ 
at the sonic point. 
Indeed, if ${B'^2}/{4\pi\rho c^2}\ll1$ then not only  
$\Gamma\approx 1$ but also the magnetic energy per particle is 
is much less than its rest mass, which means that only a small increase 
of the Lorentz factor is possible when this magnetic energy 
is converted into the kinetic one downstream. Thus, 
large asymptotic Lorentz factor implies  magnetically-dominated 
flow at the sonic point. 

In the subsonic regime different fluid elements can communicate with 
each other by means of sound waves both along and across the 
direction of motion. In the supersonic regime the causal connectivity
is limited to the interior of the Mach cone and this has an important 
implication for the efficiency of magnetic acceleration. 

%sssssssssssssssssssssssssssssssssssssssssssss
\subsection{Acceleration and differential collimation}    
%sssssssssssssssssssssssssssssssssssssssssssss

Consider first the thermal acceleration of relativistic flows 
in the supersonic regime. The mass and energy 
conservation laws for steady-state flows imply that both the mass 
and the total energy fluxes are constant along the jet: 
\begin{equation} 
\label{eq:mass-hydro}
  \rho \Gamma c A = \mbox{\small const},
\end{equation}
\begin{equation} 
\label{eq:energy-hydro}
  (\rho c^2 + 4p) \Gamma^2 c A = \mbox{\small const},
\end{equation}
where $A\propto R_j^2$ is the jet cross section area, 
$R_j$ is the jet radius, jet speed $v\approx c$, and for simplicity we
assume $\gamma=4/3$. From these equations it follows that  
\begin{equation}
\label{eq:bernoulli-hydro}
  (1 + 4p/\rho c^2) \Gamma = \Gamma_{\rm max}
\end{equation}
(this is known as the Bernoulli equation). 
$\Gamma_{\rm max}$ is the constant that equals to the Lorentz 
factor of the flow after complete conversion of its thermal
energy into the kinetic one. From the mass conservation 
we find that $\rho \propto \Gamma^{-1} R_j^{-2} $ and thus  
\begin{equation}
  p/\rho \propto \rho^{1/3} \propto \Gamma^{-1/3} R_j^{-2/3}.
\end{equation}
When $p\gg \rho$, allowing plenty of thermal energy to be spent on  
continued plasma acceleration, this equation and the Bernoulli equation 
yield
\begin{equation} 
\label{eq:gamma-hydro}
   \,\Gamma \propto R_j
\end{equation}
Thus, the sideways (or transverse) jet expansion is followed by rapid acceleration. 
In particular, for freely expanding conical jets $\Gamma\propto z$ there 
$z$ is the distance along the jet.    

For cold magnetised flows the energy equation can be written as 
\begin{equation} 
\label{eq:emergy-magnetic}
  (\rho c^2 + B'^2/4\pi) \Gamma^2 c A = \mbox{\small const}, 
\end{equation}
where we ignore the contribution due to the small poloidal component
of the magnetic field\footnote{This is sufficiently accurate when 
$R_j\gg R_{\rm LC}$, the light cylinder radius, the condition which is
normally satisfied in the supersonic regime.}. The Bernoulli equation 
then reads 
\begin{equation} 
\label{eq:bernoulli-magnetic}
  (1 + B'^2/4\pi \rho c^2) \Gamma = \Gamma_{\rm max},
\end{equation}
where $\Gamma_{\rm max}$ is the Lorentz factor after complete 
conversion of the magnetic energy into the kinetic one.
Assuming that the radius of all streamlines evolves like $R_j$ 
and using the magnetic flux freezing condition one finds the 
familiar law for the evolution of transverse magnetic field 
$B \propto R_j^{-1}$ and  $B'=B/\Gamma \propto \Gamma^{-1}R_j^{-1}$. 
This gives us  $B'^2/\rho = R_j^0 \Gamma^{-1}$ and then the 
Bernoulli equation yields the uncomfortable result  
\begin{equation} 
\label{eq:gamma-magnetic0}
   \Gamma = \mbox{\small const}. 
\end{equation}
The same conclusion can be reached in a slightly different way. 
When a fluid element expands only sideways, its volume grows as 
$V\propto R_j^{2}$ and its magnetic 
energy $e_m\propto B^2V\propto R_j^0$ remains unchanged, implying no 
conversion of the magnetic energy and no acceleration. 
Thus, in contrast to the thermal case, the sideways expansion of a 
cold magnetised flow is not sufficient for its acceleration. 

\begin{figure}[h]
\includegraphics[width=6cm]{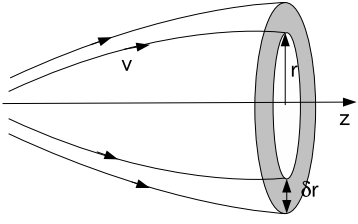}
\caption{\footnotesize Flow surfaces of steady-state axisymmetric jet.}
\label{fig:kom1}
\end{figure}

For the magnetic mechanism to work a special condition, which can 
be described as {\it differential collimation}, has to be 
satisfied. In order to see this we refine our analysis and 
consider the flow  between two axisymmetric flow surfaces 
with cylindrical radii $r(z)$ and $r(z)+\delta r(z)$ (Fig.\ref{fig:kom1}).   

\noindent
Now  $A \propto r \delta r$,  
$\rho \propto \Gamma^{-1} (r \delta r)^{-1}$, 
$B \propto \delta r^{-1}$, 
$B'^2/\rho \propto  (r/\delta r)\, \Gamma^{-1}$
and the magnetic Bernoulli equation yields
\begin{equation} 
\label{eq:gamma-magnetic}
   \Gamma \simeq \Gamma_{\rm max} 
    \left(1 - \frac{r}{\delta r}\frac{\delta r_0}{r_0} \right), 
\end{equation}
where $r_0$ and $\delta r_0$ are the initial surface parameters and 
we assume that the initial Lorentz factor $\Gamma_0\ll \Gamma_{\rm max}$. 
This result shows that the magnetic acceleration requires 
$r/\delta r$ to decrease with the distance along the jet.
In other words, the separation between neighbouring flow surfaces 
should increase faster than their radius. For example, consider a
flow with parabolic flow surfaces,  $z = z_0 (r/r_0)^a$, where 
the power index varies from surface to surface, $a=a(r_0)$.    
Then 
\begin{equation} 
  \frac{r}{\delta r} = 
  \frac{r_0}{\delta r_0} \left[
   1-\oder{a}{r_0} \frac{r_0}{a^2} \ln\frac{z}{z_0}
   \right]^{-1}. 
\end{equation}
Thus, if $da/dr_0=0$, and hence all flow surfaces are 
``uniformly'' collimated,  then $r/\delta r=$const and the 
magnetic acceleration fails. This includes the important 
case of ballistic conical flow with radial streamlines. 
If, however, $da/dr_0<0$, and thus the inner flow surfaces 
are collimated faster compared to the outer ones, then 
$r/\delta r$ decreases along the jet and the flow accelerates.  

Whether such differential collimation can arise naturally 
depends on the details of the force balance across the jet. Such 
balance is described by the Grad-Shafranov equation \citep[e.g.][]{B09}, 
which is  notoriously difficult to solve. Only very few 
analytic or semi-analytic solutions for rather simple cases have 
been found so far \citep[e.g.][]{BKR98,BN06,VK03}. 
Recently, this issue has been studied using time-dependent numerical 
simulations (Komissarov et al. 2007,2009a,b; Tchekhovskoy et al. 2009a,b).
%\citep{KBVK07,KVKB09,KVK09,TNM09,TMN09}. 
Their results shows that the differential collimation can develop 
under the self-collimating action of magnetic hoop stress associated 
with the azimuthal magnetic field, but only in cases where efficient 
externally imposed confinement helps to keep the jet 
sufficiently narrow. For example, the asymptotic Lorentz factor of conical jets 
decreases with the opening angle and unless the angle is small enough the jet 
remains Poynting-dominated. In the case of parabolic jets, 
the opening angle decreases with distance and the acceleration continues 
until the kinetic energy becomes comparable with the magnetic one.    

When the external confinement is 
provided by the gas with the pressure distribution 
$p_{\rm ext} \propto z^{-\alpha}$, where $\alpha<2$, the jet shape is 
indeed parabolic $R_j \propto z^{\alpha/4}$ and its Lorentz 
factor grows as $\Gamma \propto R_j \propto z^{\alpha/4}$ until the energy 
equipartition is reached \citep{KVKB09,Lb09}. 
As far as the dependence on $R_j$ is concerned, this is as fast as in the 
thermal mechanism. However, as a function of $z$ the Lorentz factor grows 
slower than in the case of thermally-accelerated conical jet.    
For $\alpha>2$ the external confinement is insufficient -- the 
jets eventually develop conical  streamlines and do not accelerate efficiently 
afterwards.  Various components of the Lorentz force, the hoop stress, 
magnetic pressure, and electric force, finely balance each other. 
This is in contrast to the thermal acceleration, which remains efficient 
for jets with conical geometry.

Although the detailed analysis of this issue is rather involved, one can 
get a good grasp of it via  the causality argument. 
Indeed, the favourable differential self-collimation can only be arranged 
if flow surfaces ``know'' what  other flow surfaces do. This information
is propagated by fast magnetosonic 
waves\footnote{Alfv\'en and slow magnetosonic waves transport information 
only along the magnetic field lines.}. In subsonic regime, these waves 
can propagate in all directions and have no problem in establishing 
causal communication across the jet. In the supersonic regime they are 
confined to the Mach cone which points in the direction of motion. 
When the characteristic opening angle of the Mach cone, 
\begin{equation} 
\label{eq:mach-angle}
\sin\theta_{\rm M}=\Gamma_f c_f/\Gamma v,
\end{equation}
where $c_f$ and $\Gamma_f$ are the fast magnetosonic speed and 
the corresponding Lorentz factor respectively,  
becomes smaller than the jet opening angle, $\theta_j$, 
the communication across the jet is disrupted. Thus, the condition for 
effective magnetic acceleration is $\theta_j < \theta_{\rm M}$. 
Using equations 
\ref{eq:mach-angle},\ref{eq:cf},\ref{eq:bernoulli-magnetic}, 
one can write this condition as               
\begin{equation} 
\label{eq:acc-condition}
 \Gamma < (\Gamma_{\rm max}/\sin^2\theta)^{1/3}. 
\end{equation}
For spherical wind with $\Gamma_{\rm max}\gg 1$ this condition 
reads $\Gamma<\Gamma_{\rm max}^{1/3} \ll \Gamma_{\rm max}$. 
Thus, the magnetic acceleration of relativistic winds is highly 
inefficient. Higher efficiency can be reached for collimated 
flows. For $\Gamma \ge 0.5\Gamma_{\rm max}$ 
this condition requires 
\begin{equation}
\label{eq:gamma-theta} 
  \theta_j\Gamma \le 1,
\end{equation}
where we used the small angle approximation. 
For GRB jets, with their deduced Lorentz factors as high as $\Gamma=10^3$,  
this leads to $\theta_j\le 0.06^o$. This is much less than the 
model-dependent estimates of  $2^\circ-30^\circ$ 
($\Gamma\theta_j\sim 7-70$), based on the pre-Swift observations of afterglows 
\citep{FWK00,PK01}. Moreover, 
the observed ratio of GRB and core-collapse supernova events is 
$\sim 10^{-5}$ \citep{WB06} and for beaming angles as small as $0.06^o$ we 
essentially require that every core-collapse supernova produces GRB.  
None of the current models of GRB central engines predicts such a high 
rate of GRBs and radio surveys of local SNe Ibc show that no more than 
3$\%$ of them harbour relativistic ejecta \citep{B03}. Thus, either GRB jets 
remain magnetically-dominated all the way, or other acceleration 
mechanisms come into play. During the transition from 
confined to unconfined state, which may occur when a GRB jet 
crosses the surface of collapsing star, a rarefaction wave
moves into the jet. It produces favourable differential 
collimation, due to the fact that the outer flow surfaces straighten 
up earlier compared to the inner ones, and the flow experiences 
additional acceleration \citep{KVK09,TNM09}. Based on the results of 
numerical simulations, one can expect an increase of $\Gamma\theta_j$ at most by 
a factor of ten, largely via increase of $\Gamma$. This shifts the 
theoretical value of $\theta_j$ closer to the observed range, but it is 
still a bit low. Moreover, asymptotically the flow remains rather highly 
magnetised, with approximate equipartition between the kinetic and magnetic energy 
at best. This makes shock dissipation rather ineffective.        

Condition~\ref{eq:gamma-theta} is satisfied by AGN jets, 
where $\mean{\theta_j\Gamma} \simeq 0.26$, with significant spread 
around this value \citep{P09}. However, the observed linear polarization 
angles (EVPA) of AGN jets seem to present another problem for the standard 
model. In approximately half of all cases, the electric field vector is normal 
to the jet direction \citep{W98}. This means that in the comoving jet 
frame the longitudinal component of magnetic field is at least comparable to 
the transverse one \citep{L05}. On the other hand, the standard model predicts 
that beyond the light cylinder (LC), $B_\phi/B_p \simeq (R_j/R_{\rm LC})$,
where  $B_p$ is the poloidal (predominantly longitudinal) and  
$B_\phi$ is the azimuthal components of magnetic field as measured in the 
observer's frame. In the comoving jet frame this leads to 
$B'_\phi/B'_p \simeq \Gamma^{-1} R_j/R_{\rm LC}$. For a rapidly rotating 
black hole $R_{\rm LC} \simeq 4 R_g$, where $R_g=GM/c^2$ is the 
hole's gravitational radius \citep[e.g.][]{K04}.  This leads to 
\begin{equation} 
   \frac{B'_\phi}{B'_p} \simeq \frac{10^3}{\Gamma} 
   \fracb{\theta_j}{1^\circ} \fracb{l_j}{1\mbox{pc}}
   \fracb{M}{10^8M_\odot}^{-1},        
\end{equation}     
where $l_j$ is the distance from the black hole. Thus, unless 
the AGN jets are produced by very slowly rotating black hole holes, which is 
highly unlikely, the standard model is in conflict with observations.  

According to the numerical simulations the asymptotic magnetisation of jets with 
$\Gamma_{\rm max}\sim 20$, typical for AGNs, is somewhat lower compared to that of jets 
with $\Gamma_{\rm max} \sim 1000$, typical for GRBs. However, the efficiency of shock 
dissipation is still reduced.   

%sssssssssssssssssssssssssssssssssssssssssssss
\section{Alternatives to the standard model}         
%sssssssssssssssssssssssssssssssssssssssssssss

In addition to the standard model other ideas have been put 
forward, each relaxing some of the model assumptions and exploring 
the consequences. \citet{HB00} assumed that current-driven 
instabilities randomise the magnetic field, transferring 
energy from the slowly decaying transverse component, 
$B'_\perp \propto \Gamma^{-1} R_j^{-1}$, 
to the rapidly decaying  longitudinal component, 
$B'_\parallel \propto R_j^{-2}$. 
As the result, the magnetic field strength evolves as 
$B'\propto R_j^{-2}$, and $B'^2/\rho \propto R_j^{-2}\Gamma$.          
Then in the magnetically dominated regime the Bernoulli equation 
(Eq.\ref{eq:bernoulli-magnetic}) yields $\Gamma\propto R_j$. Thus, 
randomised magnetic field behaves as ultrarelativistic gas with 
$\gamma=4/3$, providing as rapid magnetic acceleration as the 
thermal mechanism.  

Such instabilities are likely to be followed by magnetic dissipation 
and plasma heating. This may also facilitate  bulk acceleration 
of jets \citep{Dre02,DS02}. In particular, heat is easily converted into 
kinetic energy during sideways expansion of the jet. 

\citet{C95} argued that the magnetic mechanism can be more efficient 
if the jet is produced not in a steady-state but in an impulsive fashion. He  
dubbed the impulsive magnetic mechanism "astrophysical plasma gun". 
Recently, this idea have been explored in the relativistic regime and the results 
look very promising \citep{GKS10,LL10,Lt10b}. The main features of the 
``relativistic plasma gun'' mechanism are nicely demonstrated in the following 
simple problem of one-dimensional expansion of highly magnetised plasma 
into vacuum. 

\begin{figure*}
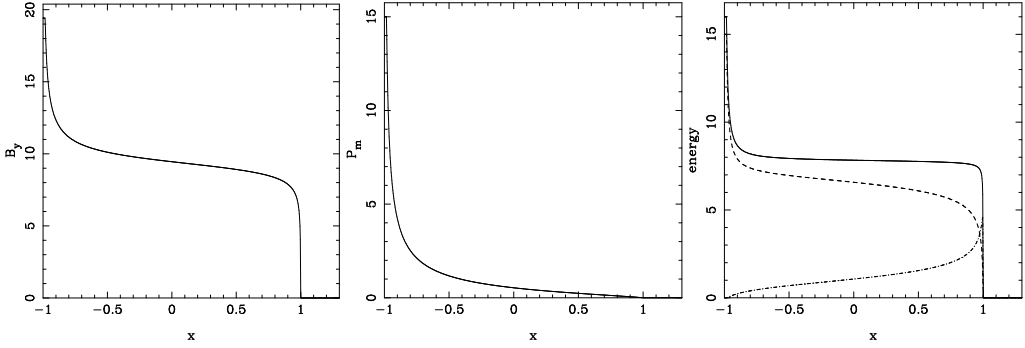

\includegraphics[width=45mm,angle=-90]{figures/byb.eps}
\includegraphics[width=45mm,angle=-90]{figures/pb.eps}
\includegraphics[width=45mm,angle=-90]{figures/t00b.eps}
\caption{ \footnotesize Solution at the time 
of separation $t=t_0$. The units are such that the dimensionless parameters 
$l_0 = 1$, $\rho_0 = 1$ and $c=1$. The wall is located at $x=-1$ and 
the initial vacuum interface is at $x=0$. The top row shows (from left 
to right) the magnetic field $B_y$, the Lorentz factor $\Gamma$, 
and the local magnetization parameter,
$\sigma=(B')^2/4\pi\rho$, where $B'$ is the magnetic field in the fluid 
frame. The bottom row shows (from left to right)
the flow velocity $v_x$, magnetic pressure, and the densities of 
total energy ({\it solid line}), magnetic energy ({\it
dashed line}), and kinetic energy ({\it dash-dotted line}) as measured
in the wall frame.  }
\label{fig:kom2}
\end{figure*}

Consider a planar uniform plasma shell of width $l_0$ with initial magnetization   
$\sigma_0 = B_0^2/4\pi\rho_0 c^2$, where $B_0$ is the initial
magnetic field and $\rho_0$ is the initial rest mass density. On the left 
of the shell is a solid conducting wall and on the right is vacuum, the magnetic 
field being parallel to the wall. When $\sigma_0\gg 1$ the particle inertia 
is very small (we assume that the plasma is cold) and has only a little 
effect on the evolution of the electromagnetic field, which closely follows the 
solution of vacuum electrodynamics. 
In the vacuum solution, an electromagnetic pulse of width $l_1 = 2l_0$ 
with constant magnetic and electric field $B=E=B_0/2$ separates from the wall 
at time $t_0\simeq l_0/c$. The corresponding exact relativistic MHD solution  
for $\sigma_0=30$ at this time is presented in figure~\ref{fig:kom2}. One 
can see that indeed $B\approx B_0/2$. However, close to the right front of the 
pulse a significant fraction 
of magnetic energy is already converted into the kinetic energy of the plasma. 
In fact, the magnetization parameter $\sigma\to 0$ at the front and 
$\Gamma \to 1+2\sigma_0$. The plasma acceleration is driven by the gradient of 
magnetic pressure (Although, in the laboratory frame the magnetic field 
is almost uniform the magnetic pressure is given by the strength of magnetic 
field in the comoving frame, $B'$, which is non-uniform.).

%fffffffffffffffffffffffffffffffffffffffffffffffffffffffffffffffffff
%fffffffffffffffffffffffffffffffffffffffffffffffffffffffffffffffffff  

%fffffffffffffffffffffffffffffffffffffffffffffffffffffffffffffffffff
\begin{figure*}
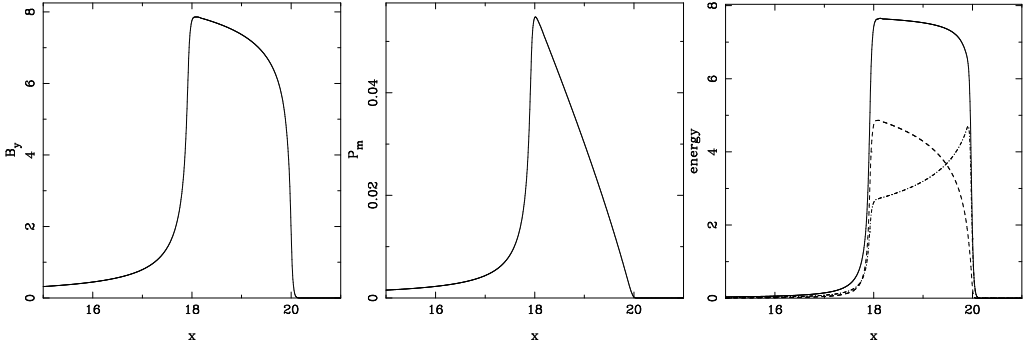
%[ht]
\includegraphics[width=45mm,angle=-90]{figures/bya.eps}
\includegraphics[width=45mm,angle=-90]{figures/pa.eps}
\includegraphics[width=45mm,angle=-90]{figures/t00a.eps}
\caption{\footnotesize The same as in Fig.~\ref{fig:kom2} but at 
$t=20t_0$.} 
\label{fig:kom3}
\end{figure*}
%fffffffffffffffffffffffffffffffffffffffffffffffffffffffffffffffffff

After the separation, a secondary rarefaction wave begins to move 
inside the pulse from the left (In Fig.\ref{fig:kom3} , which 
shows the solution at $t=20t_0$, its front is located at $x=18$.) and  
the pulse sheds plasma into the low density tail. 
However, this process is very slow and to first approximation the pulse 
rest mass, as well as its total energy and momentum, which are mainly in the 
electromagnetic form, are constant.  
However, the shell plasma continues to be accelerated 
by the magnetic pressure gradient that has developed before the 
separation. 
The rate at which the electromagnetic energy-momentum is transferred to 
plasma is dictated by the rate of the longitudinal expansion of the shell,
which is given by 
\begin{equation}
\label{ssk1}
   \oder{l}{t} = v_h - v \simeq \frac{c}{2\Gamma^2}\ , 
\end{equation}
where $v_h$ is the constant speed of the vacuum interface and 
$v$ is the characteristic (mean) speed of the shell plasma, and in 
the approximation we assume that $\Gamma_h\gg\Gamma\gg 1$. 
The electromagnetic energy of the shell,  
$ \E_{m} \simeq l B^2/4\pi \simeq \E_{m,0} (l_1/l)$,
and its kinetic energy 
$   \E_{k} = \E_{m,0} - \E_{m} \simeq \E_{m,0}( 1-(1+X)^{-1})$,  
where $X= (l-l_1)/l_1$ and $\E_{m,0}$ is the initial magnetic energy.
As long as the electromagnetic energy dominates we have $X\ll 1$ and 
can use the approximation $\E_{k} \simeq \E_{m,0} X$. On the other hand 
$\E_{k} \simeq Mc^2 \Gamma$, 
where $M=\rho \Gamma l$ is the shell rest mass. Since the secondary rarefaction, which 
develops at the back of the shell, crosses the shell very slowly, one may assume 
that $M$ is constant. Combining the last two equations we find that 
$X \simeq M\Gamma/\E_{m,0}$ and this allows us to write Eq.\ref{ssk1} as 
\begin{equation}
\label{ssk7}
  \Gamma^2 \oder{\Gamma}{t} = a\ , 
\end{equation}
where $a=\E_{m,0}c/2Ml_1=\sigma_0/8 t_0$. Integrating this equation we 
find that for $t \gg t_0$    
\begin{equation}
\label{ssk9}
  \Gamma \simeq  \sigma_0^{1/3} \fracb{t}{t_0}^{1/3},   
\end{equation}
where we ignored the factor of order unity, which is justified given 
the approximate nature of our calculations. The corresponding evolution 
of the shell thickness 
\begin{equation}
\label{ssk10}
  l \sim  l_1 \left(1+\sigma_0^{-2/3} \fracb{t}{t_0}^{1/3}\right).
\end{equation}
(This is what the shell thickness would be as the 
result of the expansion of plasma inside the shell. The secondary rarefaction reduces the 
shell thickness below this estimate.)  The condition $X\ll 1$ is no longer satisfied 
when $t$ exceeds $t_c\equiv \sigma_0^2 t_0$ as for $t=t_c$  Eq.\ref{ssk10} gives $l=2l_1$. 
At this point the shell evolution changes. Formal application of Eq.\ref{ssk9} gives the 
Lorentz factor $\Gamma=\sigma_0$ at $t=t_c$. This is larger than the value corresponding to   
full conversion of electromagnetic energy, $\Gamma_c \simeq \E_{m,0}/Mc^2 = \sigma_0/2$. 
This shows that around time $t_c$ 
the growth of Lorentz factor begins to saturate and the shell enters the coasting 
phase. Since $\Gamma_c$ is still significantly lower compared to the Lorentz factor 
of the leading front of the shell, $\Gamma_h\sim 2\sigma_0$, the shell thickness is now 
governed by the equation 
\begin{equation} 
\label{ssk12}
   \oder{l}{t} = \frac{c}{2\Gamma_c^2} \ .
\end{equation}
Integrating this equation and applying the initial condition 
$l(t_c)=2l_1$ we find 
$$
   l = l_1\left(2+\frac{t-t_c}{t_c}\right),\,\,
   \E_{m} = \E_{m,0} \left(2+\frac{t-t_c}{t_c}\right)^{-1}.
$$
For $t\gg t_c$ these equations give 
\begin{equation}
\label{ssk15}
   \frac{l}{l_0} \simeq 2\fracb{t}{t_c},\, 
   \frac{\E_{m}}{\E_{m,0}} \simeq \fracb{t}{t_c}^{-1},\,  
   \sigma\simeq \fracb{t}{t_c}^{-1}.
\end{equation}
Thus, at $t\sim 10 t_c$ essentially all electromagnetic energy is 
converted into the kinetic energy of plasma.  If the jet production is indeed highly 
intermitted and the separation between different shells is significantly larger 
than their thickness then the shock dissipation during the coasting regime can be very 
effective. Not only the flow magnetization becomes very low but the variation of 
the Lorentz factor $\Delta\Gamma\sim\Gamma$, allowing dissipation of a significant fraction
of kinetic energy.

%%%%%%%%%%%%%%%%%%%%%%%%%%%%%%%%%%%%%
\section{Discussion and Conclusions}
%%%%%%%%%%%%%%%%%%%%%%%%%%%%%%%%%%%%%
 
It appears that the properties and the potential of the magnetic mechanism 
within the framework of the standard model, which deals with 
steady-state axisymmetric ideal 
MHD flows, is now well understood. This mechanism is not as fast and 
robust as the thermal mechanism. In order to be efficient, it requires 
external confinement, which has to ensure that the  
jet remains sufficiently narrow to be causally connected in the transverse 
direction. When the external pressure distribution is a power law, 
$p_{\rm ext} \propto z^{-\alpha}$, the power index has to be below 
$\alpha_c=2$. Conical jets, which arise when such confining medium is 
not present, is an example where the standard model may fail.                

The fact that the standard model has been in the focus of theoretical 
studies for so many years is merely a reflection of its relative  
simplicity. Such issues as the jet stability, variability of central engine, and 
inhomogeneity of external medium have always been in the back mind of 
researches but it was assumed that these are details that can be considered 
later on, when the key issue of acceleration would have been settled. 
Now it appears that the standard model could be an oversimplification. 
The fine balance 
of forces in this model, which leads to the reduced efficiency of magnetic 
acceleration, may not be representative of the magnetic mechanism in general.
Instead, it may be specific to the standard model, reflecting its strict 
symmetries. As we have seen, both    
randomization of magnetic field, which is a natural outcome of magnetic 
instabilities, and impulsive operation of the jet engine, are actually 
capable of increasing effectiveness and robustness of the magnetic mechanism. 
Moreover, the observations of both AGN and GRB jets seem to require a less 
restrictive model, both in terms of polarization and jet opening angle.

%%%%%%%%%%%%%%%%%%%%%%%%%%%%%%%%%%%%%
\begin{acknowledgements} 
I am grateful to Y.Lyubarsky, J.Granot, M.Lyutikov and V.Beskin 
for their useful comments. 
\end{acknowledgements}
%%%%%%%%%%%%%%%%%%%%%%%%%%%%%%%%%%%%%

\bibliographystyle{aa}

\end{document}